\documentclass[onecolumn,showpacs]{revtex4}

\usepackage{graphicx}% Include figure files
\usepackage{dcolumn}% Align table columns on decimal point
\usepackage{bm}% bold math

\input epsf

\begin{document}

\title{\Large Role of Brans-Dicke Theory with or without self-interacting potential in cosmic acceleration}

\author{\bf Writambhara Chakraborty$^1$\footnote{writam1@yahoo.co.in}
and Ujjal Debnath$^2$\footnote{ujjaldebnath@yahoo.com} }

\affiliation{$^1$Department of Mathematics, New Alipore College, L Block, Kolkata- 700 053, India.\\
$^2$Department of Mathematics, Bengal Engineering and Science
University, Shibpur, Howrah-711 103, India. }

\date{\today}

\begin{abstract}
In this work we have studied the possibility of obtaining cosmic
acceleration in  Brans-Dicke theory with varying or constant
$\omega$ (Brans- Dicke parameter) and with or without
self-interacting potential, the background fluid being barotropic
fluid or Generalized Chaplygin Gas. Here we take the power law
form of the scale factor and the scalar field. We show that
accelerated expansion can also be achieved for high values of
$\omega$ for closed Universe.
\end{abstract}


\maketitle

\section{\normalsize\bf{Introduction}}

Recent measurements of redshift and luminosity-distance relations
of type Ia Supernovae indicate that the expansion of the Universe
is accelerating [1, 2]. This observation gives rise to the search
for a matter field, which can be responsible for accelerated
expansion. There are several proposals regarding this, {\it
Cosmological Constant, Quintessence, Dark Energy} [3 - 5] being
some of the competent candidates. However, most of these models
fit only to spatially flat $(k=0)$ Friedmann-Robertson-Walker
model [6], though a few models [7] work for open Universe
($k=-1$) also. Brans- Dicke (BD) theory has been proved to be very
effective regarding the recent study of cosmic acceleration [8].
BD theory is explained by a scalar function $\phi$ and a constant
coupling constant $\omega$, often known as the BD parameter. This
can be obtained from general theory of relativity (GR) by letting
$\omega \rightarrow \infty$ and $\phi=constant$ [9]. This theory
has very effectively solved the problems of inflation and the
early and the late time behaviour of the Universe. N. Banerjee
and D. Pavon [8] have shown that BD scalar tensor theory can
potentially solve the quintessence problem. The generalized BD
theory [10] is an extension of the original BD theory with a time
dependent coupling function $\omega$. In Generalized BD theory,
the BD parameter $\omega$ is a function of the scalar field
$\phi$. N. Banerjee and D. Pavon have shown that the generalized
BD theory can give rise to a decelerating radiation model  where
the big-bang nucleosynthesis scenario is not adversely affected
[8]. Modified BD theory with a self-interacting potential have
also been introduced in this regard. Bertolami and Martins [11]
have used this theory to present an accelerated Universe for
spatially flat model. All these theories conclude that $\omega$
should have a low negative value in order to solve the cosmic
acceleration problem. This contradicts the solar system
experimental bound $\omega\geq500$.  However Bertolami and
Martins [11] have obtained the solution for accelerated expansion
with a potential ${\phi}^{2}$ and large $|\omega|$, although they
have not considered the positive energy conditions for
the matter and scalar field.\\

In this paper, we investigate the possibilities of obtaining
accelerated expansion of the Universe in BD theory where we have
considered a self-interacting potential $V$ which is a function
of the BD scalar field $\phi$ itself and a variable BD parameter
which is also a function of $\phi$. We show all the cases of
$\omega=constant$, $\omega=\omega(\phi)$, $V=0$ and $V=V(\phi)$ to
consider all the possible solutions. We examine these solutions
for both barotropic fluid and the Generalized Chaplygin Gas [12 -
13], to get a generalized view of the results in the later case.
We analyze the conditions under which we get a negative $q$ (
deceleration parameter, $-\frac{a \ddot{a}}{{\dot{a}^{2}}}$) in
all the models of the Universe. For this purpose we have shown
the graphical representations of these scenario for further discussion.\\

The paper is organized as follows:  In section II, the field
equations for self-interacting BD theory have been given. Sections
III and IV deals with the different cases of barotropic fluid and
Generalized Chaplygin Gas respectively. Each of these two
sections are divided into two parts, namely {\bf A}($V=0$) and
{\bf B} ($V=V(\phi)$) where again two different cases have been
considered with $\omega=$ constant and $\omega=\omega(\phi)$
respectively. We have taken some particular values of the
constants for the graphical representations of $V$ and $\omega$
against the variation of the scalar field $\phi$. We have
discussed the results obtained in section V.\\

\section{\normalsize\bf{Field Equations}}

The self-interacting Brans-Dicke theory is described by the
action: (choosing $8\pi G_{0}=c=1$)

\begin{equation}
S=\int d^{4} x \sqrt{-g}\left[\phi R- \frac{\omega(\phi)}{\phi}
{\phi}^{,\alpha} {\phi,}_{\alpha}-V(\phi)+ {\cal L}_{m}\right]
\end{equation}

where $V(\phi)$ is the self-interacting potential for the BD
scalar field $\phi$ and $\omega(\phi)$ is modified version of the
BD parameter which is a function of $\phi$ [9]. The matter content
of the Universe is composed of perfect fluid,
\begin{equation}
T_{\mu \nu}=(\rho+p)u_{\mu} u_{\nu}+p~g_{\mu \nu}
\end{equation}

where $u_{\mu}~u^{\nu}=-1$ and $\rho,~p$ are respectively energy
density and isotropic pressure.\\

From the Lagrangian density $(1)$ we obtain the field equations
\begin{equation}
G_{\mu \nu}=\frac{\omega(\phi)}{{\phi}^{2}}\left[\phi  _{ , \mu}
\phi _{, \nu} - \frac{1}{2}g_{\mu \nu} \phi _{, \alpha} \phi ^{ ,
\alpha} \right] +\frac{1}{\phi}\left[\phi  _{, \mu ; \nu} -g_{\mu
\nu}~ ^{\fbox{}}~ \phi \right]-\frac{V(\phi)}{2 \phi} g_{\mu
\nu}+\frac{1}{\phi}T_{\mu \nu}
\end{equation}
and
\begin{equation}
^{\fbox{}}~\phi=\frac{1}{3+2\omega(\phi)}T-\frac{1}{3+2\omega(\phi)}\left[2V(\phi)-\phi
 \frac{dV(\phi)}{d\phi}\right]-\frac{\frac{d\omega(\phi)}{d\phi}}{3+2\omega(\phi)}{\phi,}_{\mu}
 {\phi}^{,\mu}
 \end{equation}

where $T=T_{\mu \nu}g^{\mu \nu}$.\\

The line element for Friedman-Robertson-Walker spacetime is given
by
\begin{equation}
ds^{2}=-dt^{2}+a^{2}(t)\left[\frac{dr^{2}}{1-kr^{2}}+r^{2}(d\theta^{2}+sin^{2}\theta
d\phi^{2})\right]
\end{equation}
where, $a(t)$ is the scale factor and $k(=0, \pm 1)$ is the
curvature index.\\

The Einstein field equations for the metric $(5)$ and the wave
equation for the BD scalar field $\phi$ are the following
\begin{equation}
3\frac{\dot{a}^{2}+k}{a^{2}}=\frac{\rho}{\phi}-3\frac{\dot{a}}{a}\frac{\dot{\phi}}{\phi}+\frac{\omega}{2}\frac{\dot{\phi}^{2}}{\phi
^{2}}+\frac{V(\phi)}{2\phi}
 \end{equation}
 \begin{equation}
 2\frac{\ddot{a}}{a}+\frac{\dot{a}^{2}+k}{a^{2}}=-\frac{p}{\phi}-\frac{\omega}{2}\frac{\dot{\phi}^{2}}{\phi
^{2}}-2\frac{\dot{a}}{a}\frac{\dot{\phi}}{\phi}-\frac{\ddot{\phi}}{\phi}+\frac{V(\phi)}{2\phi}
\end{equation}
and
\begin{equation}
\ddot{\phi}+3\frac{\dot{a}}{a}
\dot{\phi}=\frac{\rho-3p}{3+2\omega(\phi)}+\frac{1}{3+2\omega(\phi)}\left[2V(\phi)-\phi
\frac{dV(\phi)}{d\phi}\right]-\dot{\phi}\frac{\frac{d\omega(\phi)}{dt}}{3+2\omega(\phi)}
\end{equation}

The energy conservation equation is
\begin{equation}
\dot{\rho}+3\frac{\dot{a}}{a}(\rho+p)=0
\end{equation}

Now we consider two types of fluids, first one being the
barotropic perfect fluid and the second one is Generalized
Chaplygin gas [12, 13].

\section{\normalsize\bf{Model using barotropic fluid}}

Here we consider the Universe to be filled with barotropic fluid
with EOS
\begin{equation}
p=\gamma \rho~~~~~~~~~~~~~~~~(-1<\gamma<1)
\end{equation}
The conservation equation $(9)$ yields the solution for $\rho$ as,
\begin{equation}
\rho=\rho_{0} a^{-3(\gamma+1)}
 \end{equation}
where $\rho_{0}(>0)$ is an integration constant.\\

\subsection{\normalsize\bf{Solution without potential: $V(\phi)=0$}}

{\bf{Case I}:}\\

First we choose $\omega(\phi)=\omega$=constant.\\

Now we consider power law form of the scale factor

\begin{equation}
a(t)=a_{0} t^{\alpha}     ~~~~~(\alpha \ge 1)
\end{equation}

In view of equations $(10)$ and $(11)$, the wave equation leads
to the solution for $\phi$ to be

\begin{equation}
\phi=\frac{\rho_{0}
{a_{0}}^{-3(1+\gamma)}t^{2-3\alpha(1+\gamma)}}{(2\omega+3)(1-3\alpha
\gamma) \left[2-3\alpha(1+\gamma)\right] }
\end{equation}

 For $k\neq 0$ we get from the field equations $(6)$ and $(7)$,
 the value of $\alpha=1$ and

 \begin{equation}
 (3\gamma+1)\left[\frac{\omega}{2}(\gamma-1)(3\gamma+1)-1-\frac{k}{{a_{0}}
 ^{2} }\right]=0
 \end{equation}

 We have seen that $\gamma \ne -\frac{1}{3}$ and we have
\begin{equation}
 \omega=\frac{2(1+\frac{k}{{a_{0}} ^{2}})}{(\gamma-1)(3\gamma+1)}
 \end{equation}

 Since $\omega$ must be negative for $-\frac{1}{3}<\gamma<1$, we
 have seen that for this case the deceleration parameter $q=0$,
 i.e., the universe is in a state of uniform expansion. For $k=0$,
 the field equations yield

 \begin{equation}
 \left[2-3\alpha(\gamma+1)\right]\left[2(2\alpha-1)+\omega(\gamma-1)\{2-3\alpha(\gamma+1)\}\right]=0
 \end{equation}

 From equation $(16)$ we have two possible solutions for $\alpha$:\\

 $\alpha=\frac{2}{3(\gamma+1)}$ ~~  for $-1<\gamma < -\frac{1}{3}$\\

 and~~
 $\alpha=\frac{2\left[1+\omega(1-\gamma)\right]}{\left[4+3\omega(1-\gamma^{2})\right]}$
~~ for $-\frac{1}{3}<\gamma<1$\\

 For these values of $\alpha$, we have seen that $\omega<0$ and
 the deceleration parameter $q<0$. Thus for $k=0$ with the power
 law form of the scale factor $a=a_{0} t ^{\alpha}$ it is possible
 to get the accelerated expansion of the Universe.\\

 {\bf{Case II}:}\\

 Now we choose $\omega=\omega(\phi)$ to be variable. Here we
 consider the power law form of $\phi$ as

 \begin{equation}
 \phi(t)=\phi_{0} t^{\beta}
\end{equation}

with  the power law from of $a(t)$ given by equation $(12)$.\\

Proceeding as above we get

\begin{equation}
\omega=\frac{\alpha \beta+2\alpha+\beta-\beta ^{2}}{\beta
^{2}}-\frac{1+\gamma}{\beta ^{2}} \rho_{0} {a_{0}}^{-3(1+\gamma)}
{\phi_{0}}^{\frac{3\alpha(\gamma+1)-2}{\beta}}
\phi^{-\frac{3\alpha(\gamma+1)+\beta-2}{\beta}}+\frac{2k}{{a_{0}}^{2}
\beta^{2} {\phi_{0}} ^{\frac{2(1-\alpha)}{\beta}}}
\phi^{\frac{2(1-\alpha)}{\beta}}
\end{equation}

Now for acceleration $q<0$ implies that $\alpha>1$.
Using the other equations we arrive at two different situations:\\

 $(i)$ First considering the flat Universe model, i.e., $k=0$, we
 get, $\beta=1-3\alpha$, i.e., $\beta<-2$ for
 $\gamma>\frac{1}{3}$~,  $\beta=-2\alpha$, i.e., $\beta<-2$ (as $\alpha>1$) for
 $\gamma=\frac{1}{3}$ and  $\beta=-2$ for $\gamma<\frac{1}{3}$. That is cosmic acceleration
 can be explained at all the phases of the Universe with different values of $\beta$ where $\phi=\phi_{0} t^{\beta}$\\

 $(ii)$ If we consider the non-flat model of the Universe, i.e., $k\ne 0$, we are left with two
 options. For closed model of the Universe, i.e., for $k=1$ we can
 explain cosmic acceleration for the radiation phase only and for
 that $\beta=-2\alpha$ giving $\beta<-2$ and $6\phi_{0}
 {a_{0}}^{2}=\rho_{0}$, whereas we do not get any such possibility
 for the open model of the Universe.\\

 Now preferably taking into account the recent measurements confirming the flat model of the Universe,
 if $\beta=-2$ we see that we have an accelerated expansion of
 the Universe after the radiation period preceded by a decelerated
 expansion before the radiation era and a phase of uniform
 expansion at the radiation era itself. Also if $\beta<-2$ cosmic
 acceleration is followed by a deceleration phase as $\alpha<
 1$ for $\gamma<\frac{1}{3}$.\\

\begin{figure}
\includegraphics[height=1.2in]{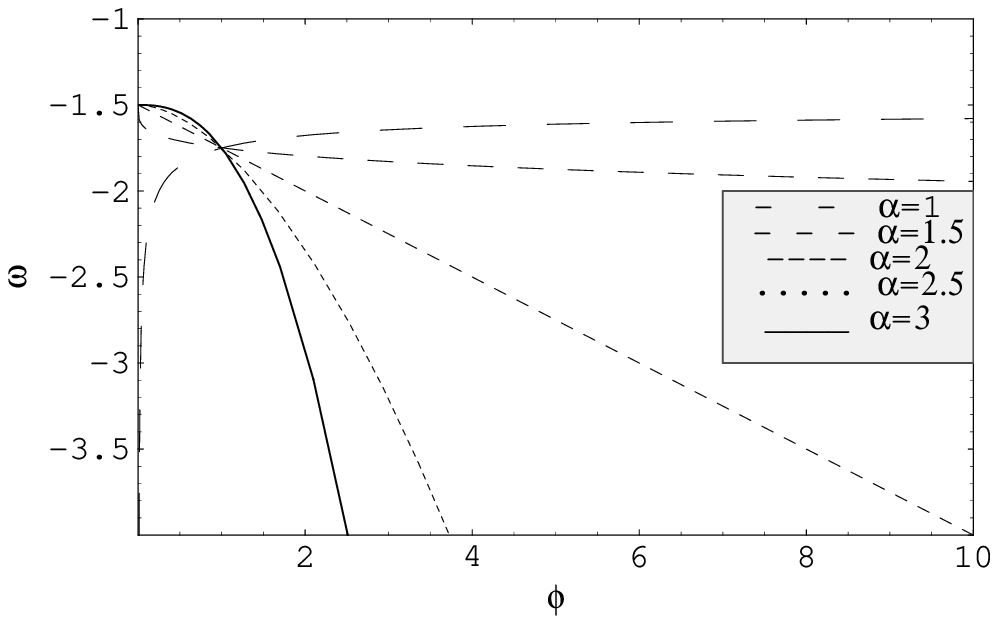}~~~~~~~~~~~~~~~~~~~~
\includegraphics[height=1.2in]{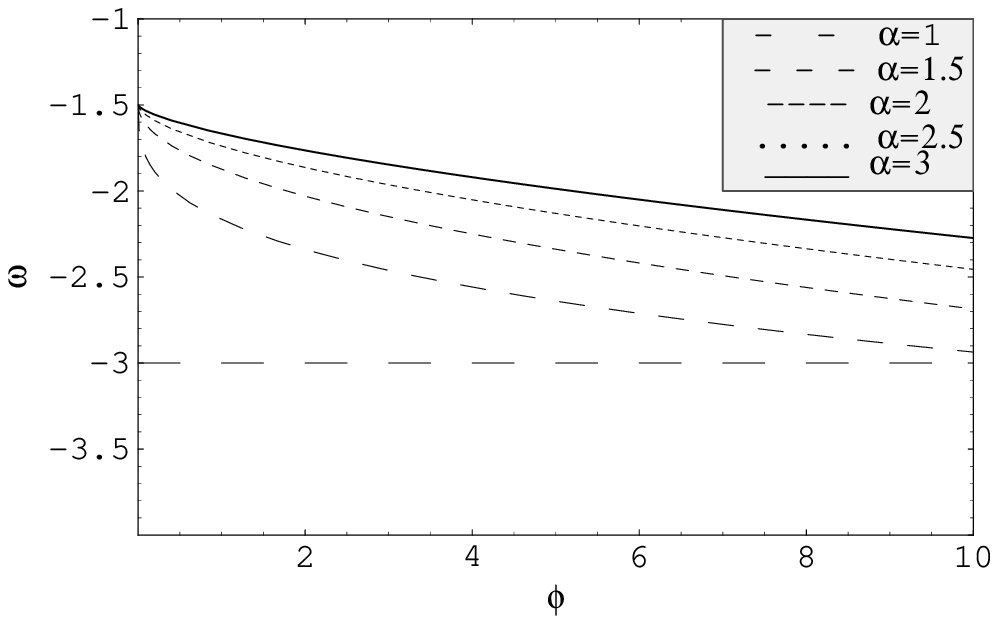}\\
\vspace{1mm}
Fig.1~~~~~~~~~~~~~~~~~~~~~~~~~~~~~~~~~~~~~~~~~~~~~~~~~~~~~~~~~~~Fig.2\\
\vspace{2mm}barotropic fluid:\\
 \vspace{2mm}
$\omega=\omega(\phi),k=0,\beta=-2(dust)$~~~~~~~~~~~~~~~~~~~~~~~~~~~~~~~~~~~~
$\omega=\omega(\phi),k=1,\beta=-2\alpha,\gamma=\frac{1}{3}$\vspace{5mm}

\vspace{5mm} Fig. 1 and 2 shows the variation of $\omega$
 against $\phi$ for different values of $\alpha=1, 1.5, 2, 2.5,3$.
 In Fig 1 we have considered
 flat model, i.e., $k=0$ and the present dust filled epoch, i.e., $\gamma=0$,
  normalizing the parameters as $a_{0}=\rho_{0}=\phi_{0}=1$,
  also as the calculation shows, for this $\beta=-2$, whereas for Fig 2 we have taken closed model, i.e., $k=1$ and
  $a_{0}=\phi_{0}=1, \rho_{0}=6, \gamma=\frac{1}{3}$ and $\beta=-2\alpha$, according to the calculations. \hspace{14cm} \vspace{4mm}

\end{figure}

\subsection{\normalsize\bf{Solution with potential: $V=V(\phi)$}}

{\bf{Case I}:}\\

Let us choose $\omega(\phi)=\omega=$constant.\\

In this case instead of considering equations $(12)$ and $(17)$ we
consider only one power law form

\begin{equation}
\phi=\phi_{0} a^{\alpha}
\end{equation}

Using equation $(19)$ in equations $(6)$ and $(7)$ we get

$$
\dot{a}={\left[2k+2(1+\gamma)\frac{\rho_{0}}{\phi_{0}}\frac{a^{-3\gamma-\alpha-1}}{\{3\gamma
\alpha+6\gamma-\alpha^{2}+7\alpha+6-2\omega \alpha^{2}\}
}\right]}^{\frac{1}{2}} $$

 Putting $k=0$, we get

\begin{equation}
a=A t^{\frac{2}{3+\alpha+3\gamma}}
\end{equation}

where $A={\left[\frac{\rho_{0}
(1+\gamma)(3+\alpha+3\gamma)^{2}}{2\rho_{0}\{6(1+\gamma)+\alpha(7+3\gamma)-\alpha^{2}(1+2\omega)\}}\right]}^{\frac{1}{3+\alpha+3\gamma}}$.\\

Therefore, $\phi=B t^{\frac{2\alpha}{3+\alpha+3\gamma}}$ where,
$B=\phi_{0} A^{\alpha}$\\

Now, if $\frac{2}{3+\alpha+3\gamma}\ge 1$, we get

\begin{equation}
\alpha\le -(1+3\gamma)
\end{equation}

Substituting these values in $(6), (7), (8)$, the solution for
the potential $V$ is obtained as,
$V=\frac{B'}{\phi^{\frac{3+3\gamma}{\alpha}}}$ where,
$B'=-\frac{2B\{6-18\alpha+6\omega \alpha+6\omega \alpha
\gamma-18\gamma}{3(3+3\gamma+\alpha)^{2}(1+\gamma)}$.\\

Also, the deceleration parameter reduces to, $q=-\frac{a
\ddot{a}}{{\dot{a}}^{2}}=\frac{3\gamma+\alpha+1}{2}\le 0$
~~~~(using equation $(21)$)\\

Hence, the present Universe is in a state of expansion with
acceleration.\\

Also, we get
$\omega=-\frac{6\gamma(1+\gamma)}{\alpha}-\frac{3+\alpha}{2\alpha}$
and, $\alpha=-\frac{3(1+2\gamma)^{2}}{1+2\omega}$.\\

Also, $\gamma\ge -1 \Rightarrow \alpha \le 2$ and $\omega \ge
-\frac{5}{4}$. For the present Universe (i.e., taking $\gamma=0$)
and the $\Lambda$CDM model,
$\omega=-\frac{3+\alpha}{2\alpha}$.\\\\

{\bf{Case II}:}\\
\begin{figure}
\includegraphics[height=1.2in]{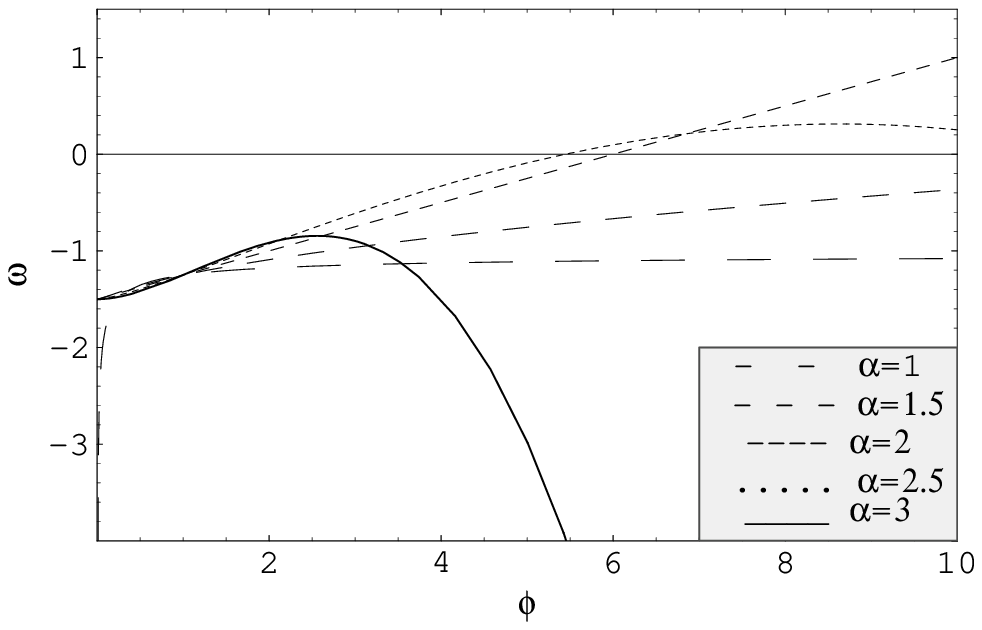}~~~~~~~~~~~~~~~~~~~
\includegraphics[height=1.2in]{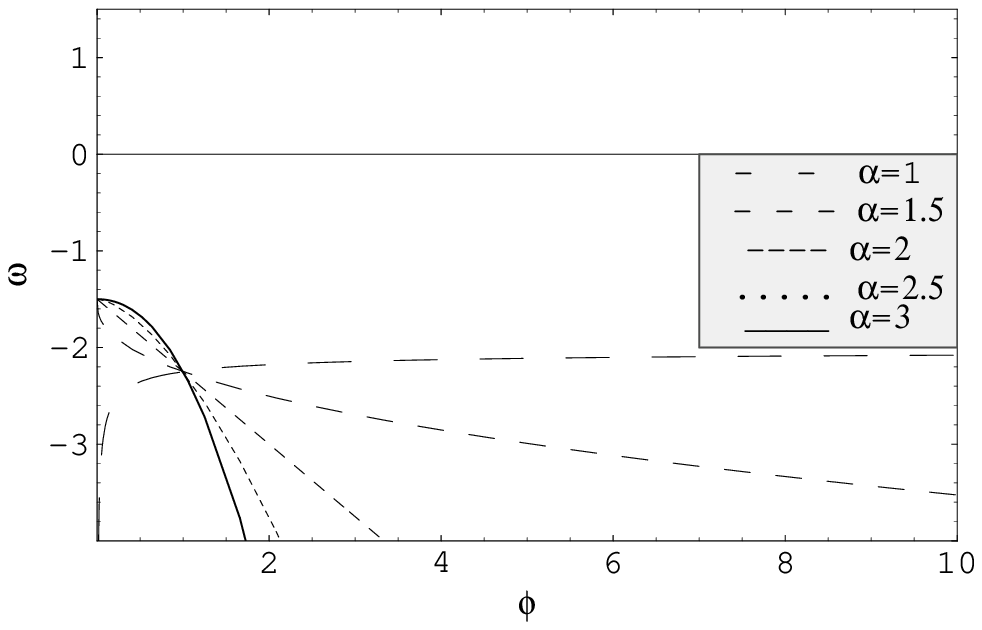}\\
\vspace{1mm}
Fig.3~~~~~~~~~~~~~~~~~~~~~~~~~~~~~~~~~~~~~~~~~~~~~~~~~~~~~Fig.4\\
\vspace{2mm}barotropic
fluid:(dust)\\
\vspace{2mm}$\omega=\omega(\phi),k=1,\beta=-2$~~~~~~~~~~~~~~~~~~~~~~~$\omega=\omega(\phi),k=-1,\beta=-2$\\
 \vspace{5mm}

\vspace{5mm} Fig. 3 and 4 shows the variation of $\omega$
 against $\phi$ for respectively closed and open models of the Universe. We take
 different values of $\alpha=1, 1.5, 2, 2.5,3$.
 In both the figures we have considered
 the present dust filled epoch, i.e., $\gamma=0$ and $\beta=-2$,
  normalizing the parameters as $a_{0}=\rho_{0}=\phi_{0}=1$.  \hspace{14cm} \vspace{4mm}

\end{figure}

Now we choose $\omega(\phi)$ to be dependent on $\phi$. Again we
consider the power law forms, $(12)$ and $(17)$. Solving the
equations in a similar manner, we get

\begin{equation}
\omega=\frac{\alpha \beta+2\alpha+\beta-\beta ^{2}}{\beta
^{2}}-\frac{1+\gamma}{\beta ^{2}} \rho_{0} {a_{0}}^{-3(1+\gamma)}
{\phi_{0}}^{\frac{3\alpha(\gamma+1)-2}{\beta}}
\phi^{-\frac{3\alpha(\gamma+1)+\beta-2}{\beta}}+\frac{2k}{{a_{0}}^{2}
\beta^{2} {\phi_{0}} ^{\frac{2(1-\alpha)}{\beta}}}
\phi^{\frac{2(1-\alpha)}{\beta}}
\end{equation}
and
\begin{equation}
V(\phi)=(2\alpha+\beta)(3\alpha+2\beta-1)
{\phi_{0}}^{\frac{2}{\beta}}
\phi^{\frac{\beta-2}{\beta}}-(1-\gamma) \rho_{0}
{a_{0}}^{-3(1+\gamma)}
{\phi_{0}}^{\frac{3\alpha(\gamma+1)}{\beta}}
\phi^{-\frac{3\alpha(\gamma+1)}{\beta}}+\frac{4k}{{a_{0}}^{2}}\phi^{\frac{\beta-2\alpha}{\beta}}{\phi_{0}}^{\frac{2\alpha}{\beta}}
\end{equation}

Substituting these values in equation $(8)$, we get

\begin{equation}
\text either~~~~ \beta=-2   ~~~~~\text or ~~~~\beta=-2\alpha
\end{equation}

Therefore for cosmic acceleration $q<0\Rightarrow \alpha>1$ and
$\beta\le -2$.\\

Therefore for the present era,

\begin{eqnarray*}
\omega=-\frac{3}{2}-\frac{\rho_{0}{a_{0}}^{-3}}{4\phi_{0}}
t^{\frac{3\alpha-4}{2}}~~~~~\text and~~~V=2(\alpha-1)(3\alpha-5)
\phi_{0} t -t^{\frac{3\alpha}{2}}\rho_{0}{a_{0}}^{-3} ~~~\text
if~~~ \beta=-2
\end{eqnarray*}
\begin{equation}
\omega=-\frac{3}{2}-\frac{\rho_{0}{a_{0}}^{-3}}{\phi_{0}}
t^{\frac{\alpha-2}{2\alpha}}~~~~\text{and} ~~~
V=-\phi_{0}{a_{0}}^{-3}t^{\frac{3}{2}}~~~\text{if} ~~~\beta<-2
\end{equation}

Also for vacuum dominated era,

\begin{eqnarray*}
\omega=-\frac{3}{2} ~~~~\text{and}~~~~ V=
2(\alpha-1)(3\alpha-5)t-2\rho_{0}~~~~\text for ~~~\beta=-2
\end{eqnarray*}
\begin{equation}
\omega=-\frac{3}{2}    ~~~\text and ~~~~V=-2\rho_{0}~~~\text for
~~~\beta<-2
\end{equation}

\begin{figure}
\includegraphics[height=1.3in]{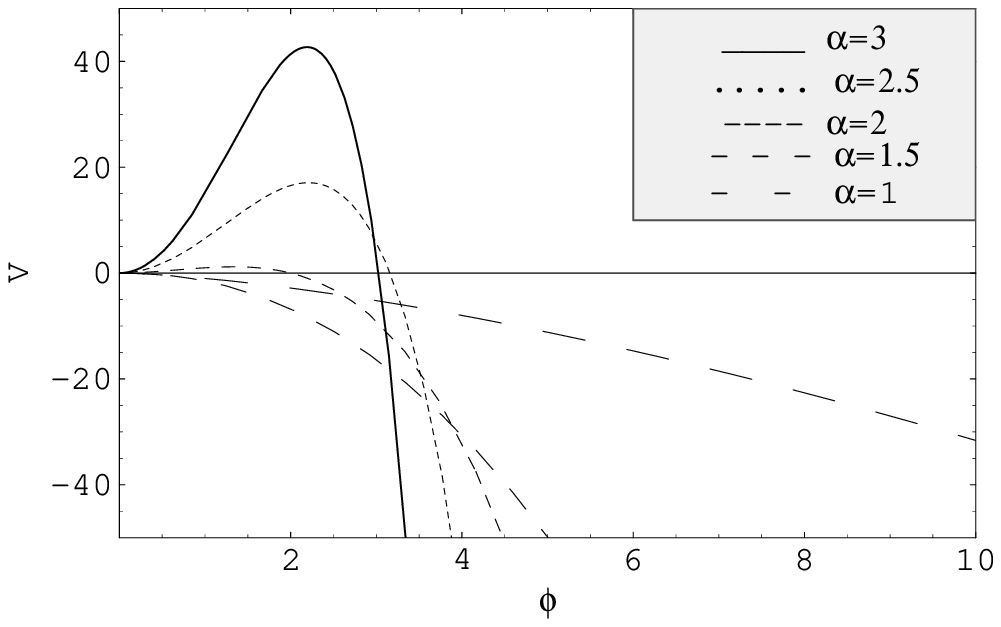}
\includegraphics[height=1.3in]{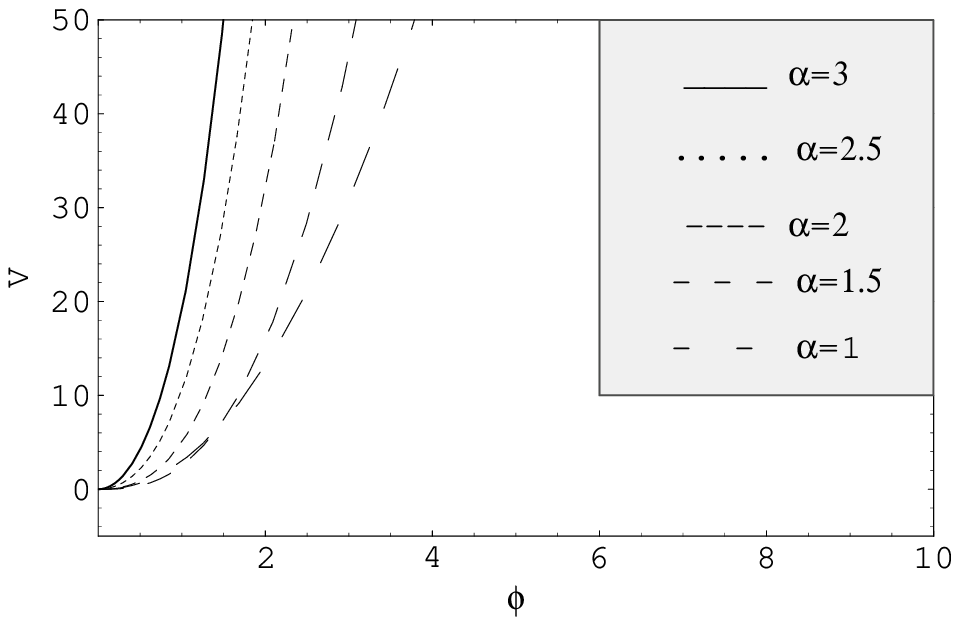}
\includegraphics[height=1.3in]{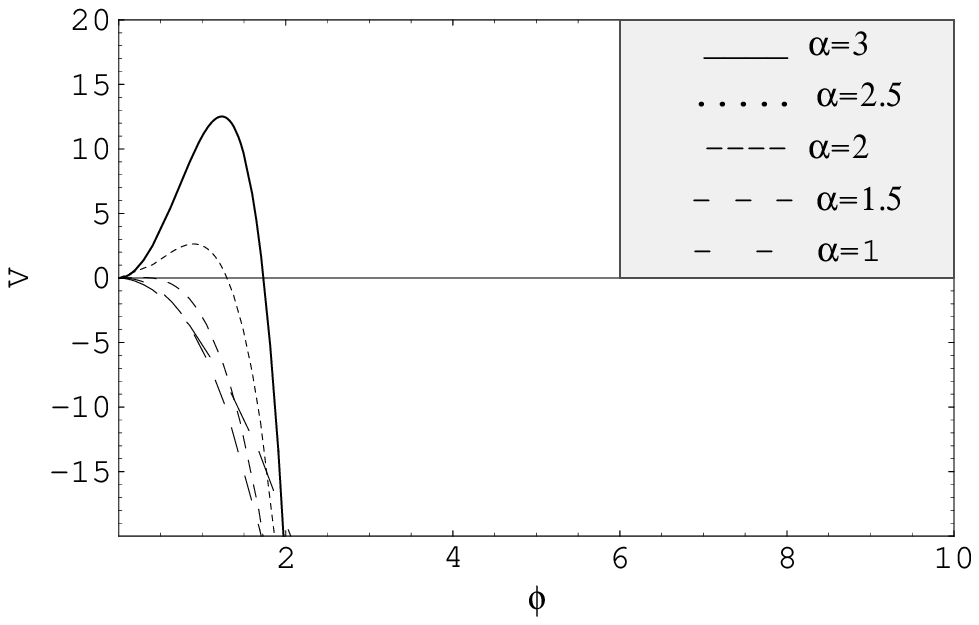}\\
\vspace{1mm}
Fig.5~~~~~~~~~~~~~~~~~~~~~~~~~~~~~~~~~~~~~~~~Fig.6~~~~~~~~~~~~~~~~~~~~~~~~~~~~~~~~~~~~~~~~~~~~~Fig.7\\
\vspace{2mm}barotropic fluid:(dust)\\
\vspace{2mm}
$V=V(\phi),\beta=-2,k=0$~~~~~~~~~~~~~~~~$V=V(\phi),\beta=-2,k=1$~~~~~~~~~~~~~~~~~~~~~~~~$V=V(\phi),\beta=-2,k=-1$
\vspace{5mm}

\vspace{5mm} Fig. 5, 6 and 7 shows the variation of $V$
 against $\phi$ for respectively flat, closed and open models of the Universe.
 We have considered different values of $\alpha=1, 1.5, 2, 2.5,3$ and $\beta=-2$.
 In all the three the figures we have considered
 the present dust filled epoch, i.e., $\gamma=0$,
  normalizing the parameters as $a_{0}=\rho_{0}=\phi_{0}=1$. \hspace{14cm} \vspace{4mm}

\end{figure}

\section{\normalsize\bf{Model using Generalized Chaplygin Gas}}

Here we consider the Universe to be filed with Generalized
Chaplygin Gas with EOS
\begin{equation}
p=-\frac{B}{\rho^{n}}
\end{equation}
Here the conservation equation $(9)$ yields the solution for
$\rho$ as,
\begin{equation}
\rho=[B+\frac{C}{a^{3(1+n)}}]^{\frac{1}{(1+n)}}
\end{equation}
where $C$ is an integration constant.

\begin{figure}
\includegraphics[height=1.2in]{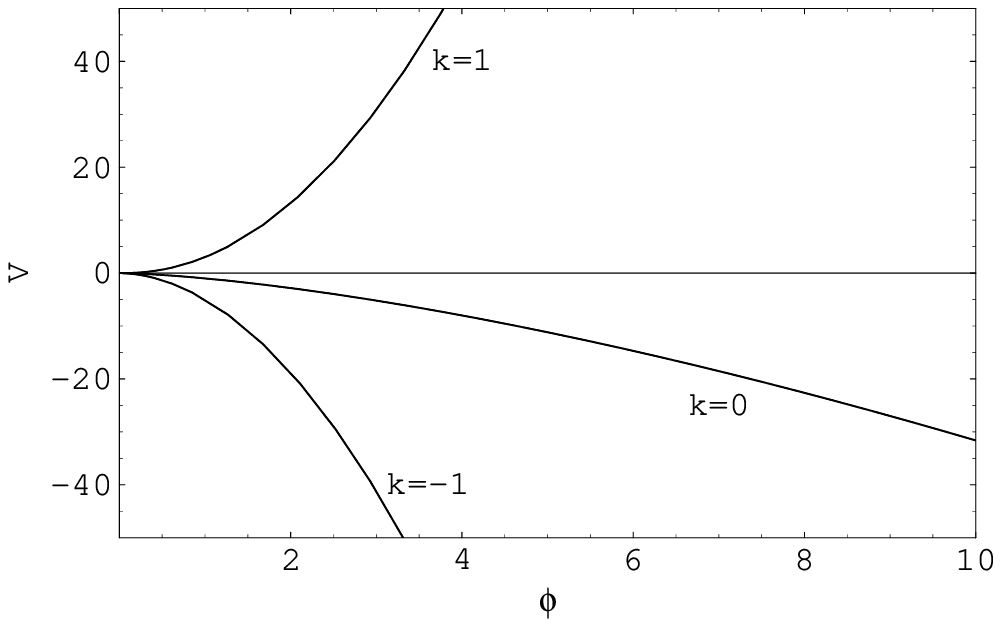}~~~~~~~~~~~~~~~~~~~~~
\includegraphics[height=1.2in]{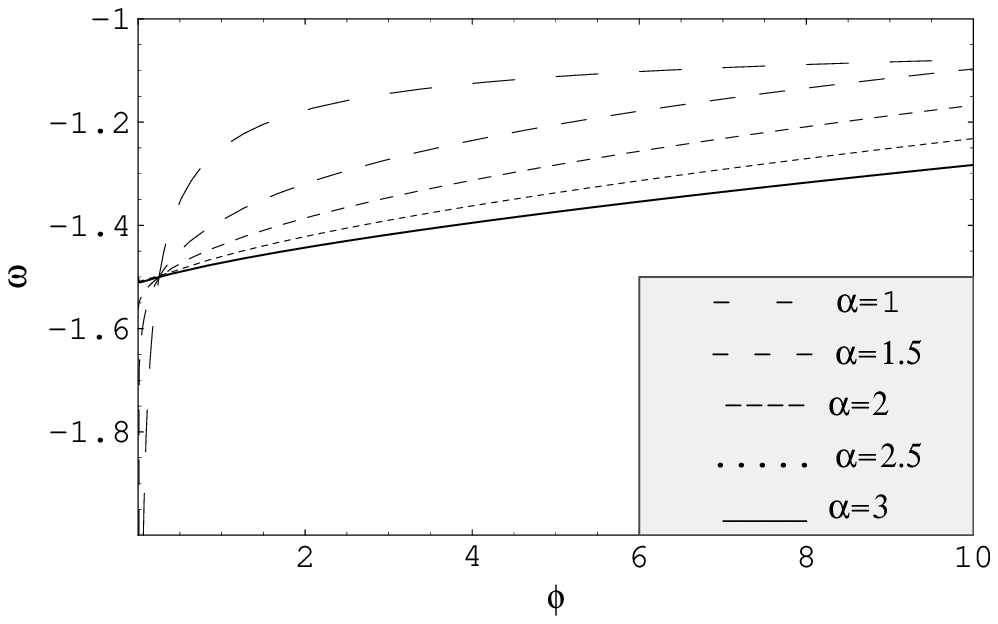}\\
\vspace{2mm}Fig.8~~~~~~~~~~~~~~~~~~~~~~~~~~~~~~~~~~~~~~~~~~~~~~~~~~~~~~~~~~~~Fig.9\\
\vspace{2mm}barotropic
fluid:(dust)\\
\vspace{2mm}$V=V(\phi),\beta=-2\alpha,\gamma=0$~~~~~~~~~~~~~~~~~~~~~~~~~~~~~~~~$\omega=\omega(\phi),k=1,\beta=-2\alpha,\gamma=0$
 \vspace{5mm}

\vspace{5mm} Fig. 8 shows the variation of $V$ against the
variation of $\phi$ for all the models of the Universe, whereas,
fig. 9 shows the variation of $\omega$ for only the closed model
of the Universe. We have considered different values of $\alpha=1,
1.5, 2, 2.5,3$ and the present dust filled epoch, i.e.,
$\gamma=0$, normalizing the parameters as
$a_{0}=\rho_{0}=\phi_{0}=1$, also as the calculation shows, for
this $\beta=-2\alpha$. For figure 8 the results for different
values of $\alpha$ coincides with each other in each model of the
Universe.\hspace{14cm} \vspace{4mm}

\end{figure}
\subsection{\normalsize\bf{Solution without potential: $V(\phi)=0$}}

{\bf{Case I}:}\\

First we choose $\omega(\phi)=\omega=$constant.\\

We consider the power law form

\begin{equation}
\phi=\phi_{0} a^{\alpha}
\end{equation}
Equations $(6), (7), (8)$ give,
\begin{equation}
(2\omega \alpha-6)\ddot{a}+(\omega \alpha^{2}+4\omega
\alpha-6)\frac{{\dot{a}}^{2}}{a}=\frac{6}{a}k
\end{equation}
which yields the solution,
\begin{equation}
\dot{a}=\sqrt{\frac{6k}{P(\omega \alpha-3)}+K_{0}a^{-P}}
\end{equation}
where $P=\frac{\omega \alpha^{2}+4\omega \alpha-6}{\omega
\alpha-3}$ and $K_{0}$ is an integration constant.\\

First we consider $P>0$. Multiplying both sides of equation $(31)$
by $a^{P}$ after squaring it, we get $K_{0}=0$, therefore giving,
$a=\sqrt{\frac{6k}{(\omega \alpha-3)P}}t$.\\

Hence for flat Universe, we get, $a=$constant.\\

For open model, we must have $\omega \alpha<3$ and
$a=\sqrt{\frac{6}{(3-\omega \alpha)P}}t$, whereas, for closed
model, $\omega \alpha>3$ and $a=\sqrt{\frac{6}{(\omega
\alpha-3)P}}t$. In all cases $q=0$, i.e., we get uniform
expansion.\\

If $P=0$, $a \ddot{a}=\frac{3}{\omega \alpha-3}k$, i.e.,
$\dot{a}^{2}=\frac{6k}{\omega \alpha-3}\ln{a}+K_{0}$.\\

If $k=0$, $a=\sqrt{K_{0}}t+C_{0}$, ($C_{0}$ is an integration
constant) causing $q=0$, i.e., uniform expansion again.\\

\begin{figure}
\includegraphics[height=1.3in]{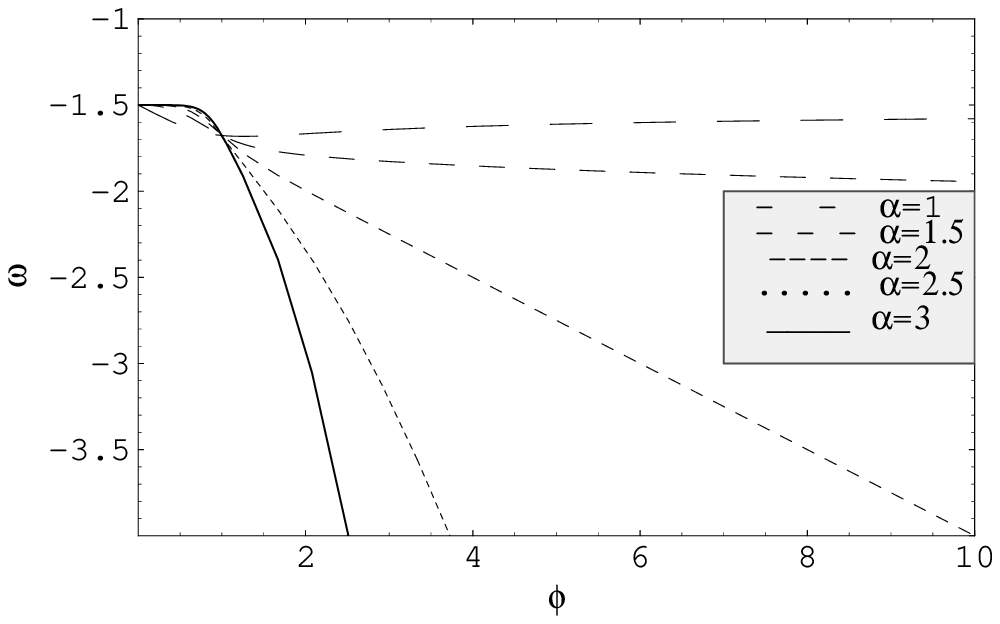}
\includegraphics[height=1.3in]{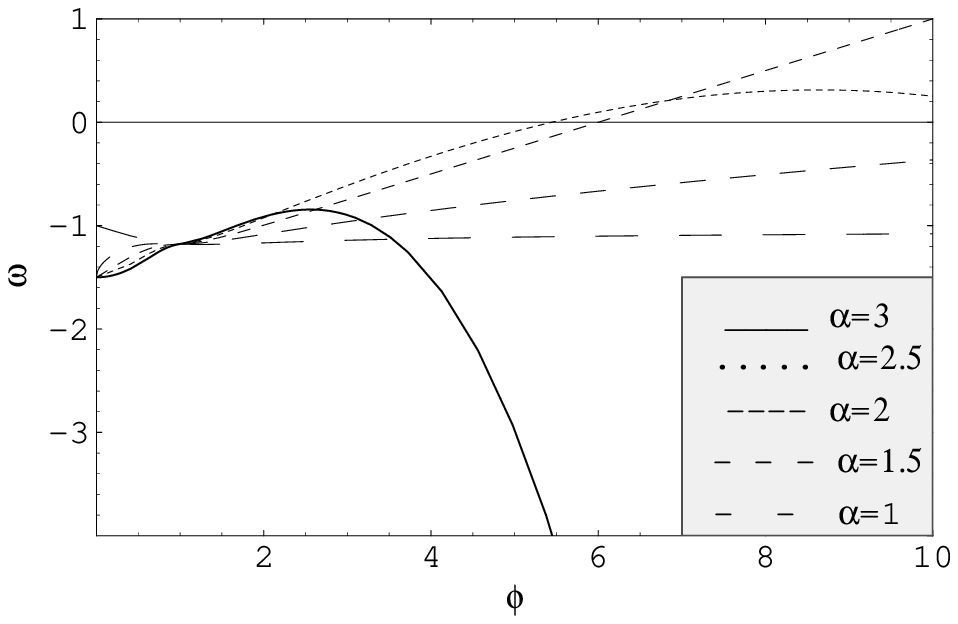}
\includegraphics[height=1.3in]{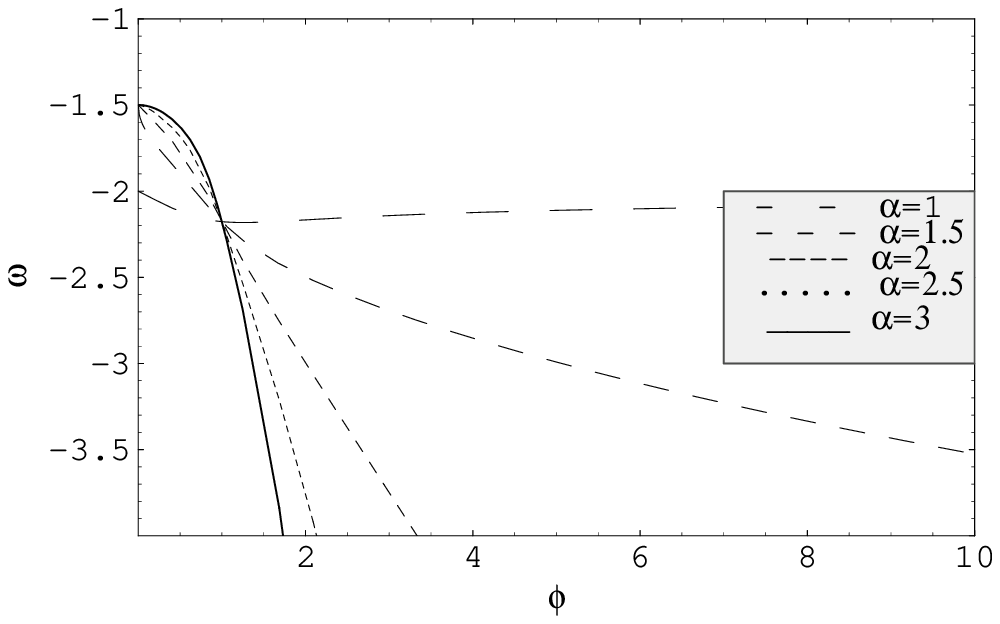}\\
\vspace{1mm}
Fig.10~~~~~~~~~~~~~~~~~~~~~~~~~~~~~~~~~~~~~~~~Fig.11~~~~~~~~~~~~~~~~~~~~~~~~~~~~~~~~~~~~~~~~~~~~~Fig.12\\
\vspace{2mm}Generalized Chaplygin Gas:\\
\vspace{2mm}
$\omega=\omega(\phi),\beta=-2,k=0$~~~~~~~~~~~~~~~~$\omega=\omega(\phi),\beta=-2,k=1$~~~~~~~~~~~~~~~~~~~~~~~~$\omega=\omega(\phi),\beta=-2,k=-1$
\vspace{5mm}

\vspace{5mm} Fig. 10, 11 and 12 shows the variation of $\omega$
 against $\phi$ for respectively flat, closed and open models of the Universe.
 We have considered different values of $\alpha=1, 1.5, 2, 2.5,3$ and $\beta=-2$.
 In all the three the figures we have considered
$n=1$,  normalizing the parameters as $a_{0}=\rho_{0}=\phi_{0}=B=C=1$. \hspace{14cm} \vspace{4mm}

\end{figure}

 {\bf{Case II}:}\\

 Now we consider $\omega=\omega(\phi)$, i.e., $\omega$ dependent on
 $\phi$.\\

 Also the power law forms considered will be $(12)$ and
 $(17)$. Solving the equations we get,
\begin{equation}
\omega(\phi)=\frac{\alpha \beta+2\alpha+\beta-\beta ^{2}}{\beta
^{2}}-\frac{C{a_{0}}^{-3(1+n)}{\phi_{0}}^{\frac{3\alpha(1+n)-2}{\beta}}
\phi^{\frac{-3\alpha(1+n)-\beta+2}{\beta}}}
{\beta^{2}\left[B +
C{a_{0}}^{-3(1+n)}{\phi_{0}}^{\frac{3\alpha(1+n)}{\beta}}
\phi^{\frac{-3\alpha(1+n)}{\beta}}\right]^{\frac{n}{1+n}}}
+\frac{2k \phi^{\frac{2(1-\alpha)}{\beta}}}{{a_{0}}^{2}
\beta^{2}{\phi_{0}}^{\frac{2(1-\alpha)}{\beta}}}
\end{equation}
Also substituting these values  in the given equations, we get,
either $n=-1$ or $B=0$ and also $k=0$. If $n=-1$, we
get back barotropic fluid, and if $B=0$, we get dust filled
Universe. In both the cases the Generalized Chaplygin gas does
not seem to have any additional effect on the cosmic
acceleration.\\

\subsection{\normalsize\bf{Solution with potential: $V=V(\phi)$}}

{\bf{Case I}:}\\

Let us choose $\omega(\phi)=\omega=$constant.\\

We again consider the power law forms $(12)$ and $(17)$. We get
the solution for $V(\phi)$ to be

\begin{equation}
V(\phi)=(2\alpha+\beta)(3\alpha+2\beta-1)
{\phi_{0}}^{\frac{2}{\beta}}
\phi^{\frac{\beta-2}{\beta}}+\frac{-2B-C{a_{0}}^{-3(1+n)}
{\phi_{0}}^{\frac{3\alpha(1+n)}{\beta}}\phi^{\frac{-3\alpha(1+n)}{\beta}}}{\left[B
+
C{a_{0}}^{-3(1+n)}{\phi_{0}}^{\frac{3\alpha(1+n)}{\beta}}
\phi^{\frac{-3\alpha(1+n)}{\beta}}\right]^{\frac{n}
{(1+n)}}}+\frac{4k}{{a_{0}}^{2}}\phi^{\frac{\beta-2\alpha}{\beta}}{\phi_{0}}^{\frac{2\alpha}{\beta}}
\end{equation}

Substituting these values in the other equations we get that
$n=-1$, i.e., the equation of state of Generalized Chaplygin Gas
takes the form of that of barotropic fluid. Also we get,
$\alpha=1$, which implies $q=0$, i.e., uniform expansion of the
Universe.\\

\begin{figure}
\includegraphics[height=1.3in]{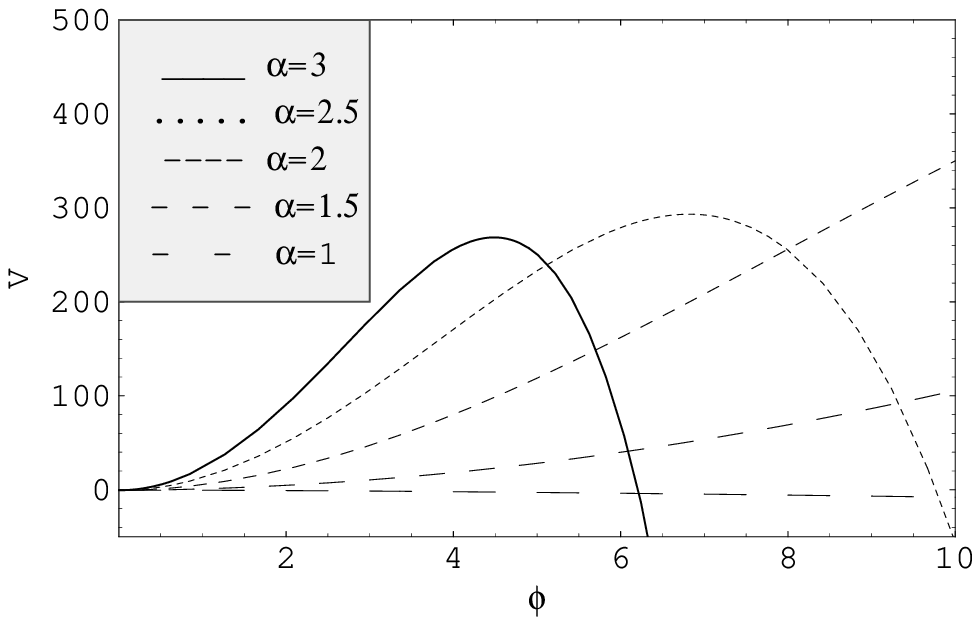}
\includegraphics[height=1.3in]{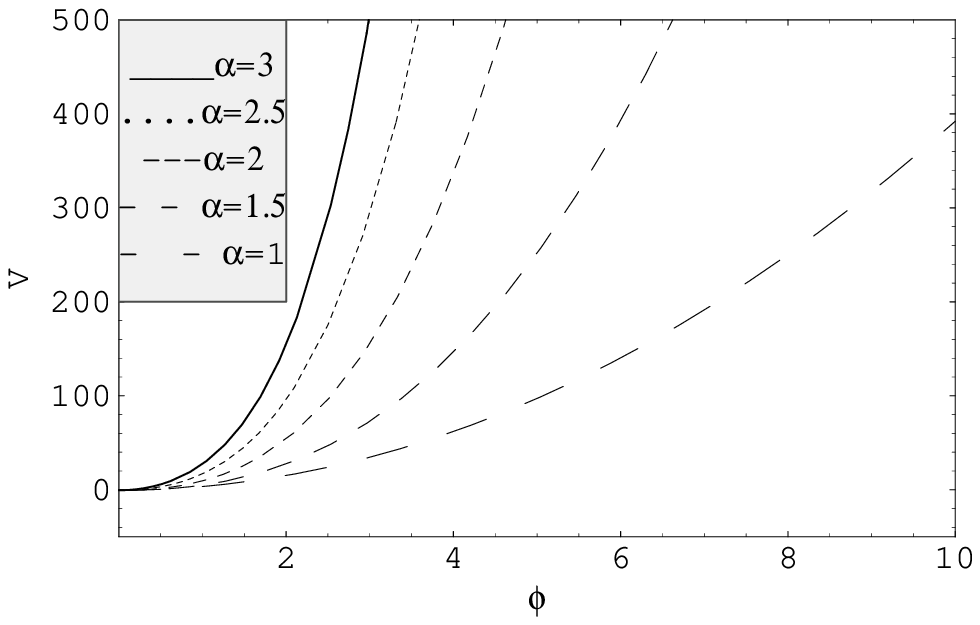}
\includegraphics[height=1.3in]{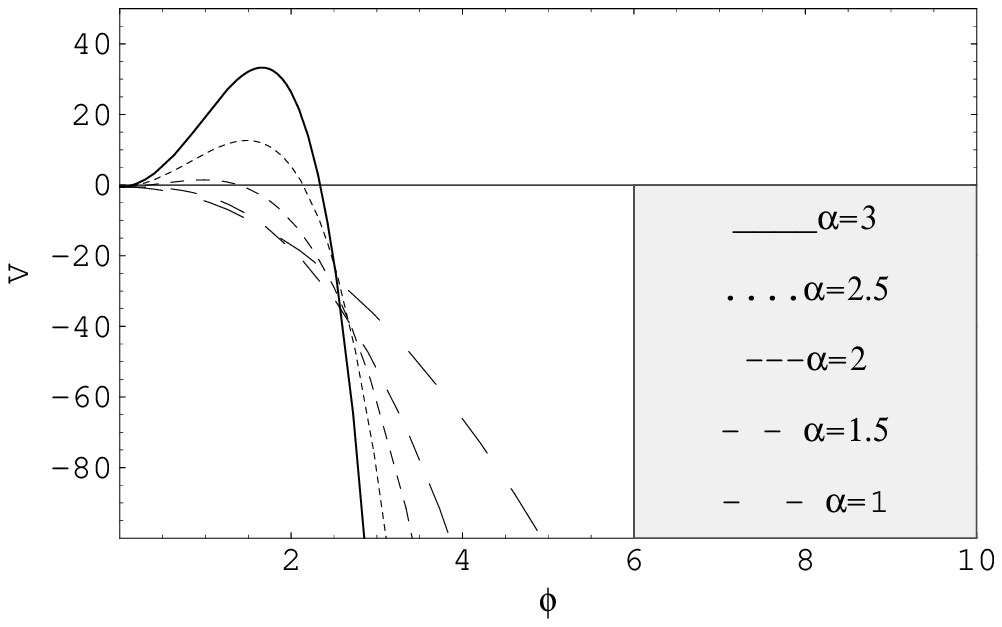}\\
\vspace{1mm}
Fig.13~~~~~~~~~~~~~~~~~~~~~~~~~~~~~~~~~~~~~~~~Fig.14~~~~~~~~~~~~~~~~~~~~~~~~~~~~~~~~~~~~~~~~~~~~~Fig.15\\
\vspace{2mm}Generalized Chaplygin Gas:\\
\vspace{2mm}
$V=V(\phi),\beta=-2,k=0$~~~~~~~~~~~~~~~~$V=V(\phi),\beta=-2,k=1$~~~~~~~~~~~~~~~~~~~~~~~~$V=V(\phi),\beta=-2,k=-1$
\vspace{5mm}

\vspace{5mm} Fig. 13, 14 and 15 shows the variation of $V$
 against $\phi$ for respectively flat, closed and open models of the Universe.
 We have considered different values of $\alpha=1, 1.5, 2, 2.5,3$ and $\beta=-2$.
 In all the three the figures we have considered
$n=1$,  normalizing the parameters as $a_{0}=\rho_{0}=\phi_{0}=B=C=1$. \hspace{14cm} \vspace{4mm}

\end{figure}

{\bf{Case II}:}\\

Now we choose $\omega(\phi)$ to be dependent on $\phi$.\\

Again we consider the power law forms, $(12)$ and $(17)$. Solving
the equations we get the solutions for Brans-Dicke parameter and
self-interacting potential as same as equations $(32)$ and $(33)$
respectively.\\

Substituting these values in equation $(8)$, we get

\begin{equation}
\text either~~~~ \beta=-2   ~~~~~\text or ~~~~\beta=-2\alpha
\end{equation}

Therefore for cosmic acceleration $q<0\Rightarrow \alpha>1$ and
$\beta\le -2$.\\

Therefore for the dust dominated era,

\begin{eqnarray*}
\omega=-\frac{3}{2}-\frac{\rho_{0}
{a_{0}}^{-3}\phi^{\frac{3\alpha-4}{2}}}{4{\phi_{0}}^{\frac{3\alpha-2}{2}}}~~~~~\text
and~~~V=\frac{2(\alpha-1)(3\alpha-5)}{\phi_{0}}\phi^{2}-\rho_{0}{a_{0}}^{-3}
\frac{\phi^{\frac{3\alpha}{2}}}{{\phi_{0}}^{\frac{3\alpha}{2}}}
~~~\text if~~~ \beta=-2
\end{eqnarray*}
\begin{equation}
\omega=-\frac{3}{2}-\frac{\rho_{0}
{a_{0}}^{-3}\phi^{\frac{\alpha-2}{2\alpha}}}{4
\alpha^{2}{\phi_{0}}^{\frac{3\alpha-2}{2\alpha}}}~~~~\text{and}
~~~
V=-\rho_{0}{a_{0}}^{-3}\frac{\phi^{\frac{3}{2}}}{{\phi_{0}}^{\frac{3}{2}}}~~~\text{if}
~~~\beta<-2
\end{equation}

Also for vacuum dominated era,

\begin{eqnarray*}
\omega=-\frac{3}{2} ~~~~\text{and}~~~~ V=
2(\alpha-1)(3\alpha-5)\frac{\phi^{2}}{\phi_{0}}-2\left[\rho_{vac}\right]_{\beta=-2}~~~~\text
for ~~~\beta=-2
\end{eqnarray*}
\begin{equation}
\omega=-\frac{3}{2}    ~~~\text and
~~~~V=-2\left[\rho_{vac}\right]_{2\alpha+\beta=0}~~~\text for
~~~\beta<-2
\end{equation}

\begin{figure}
\includegraphics[height=1.3in]{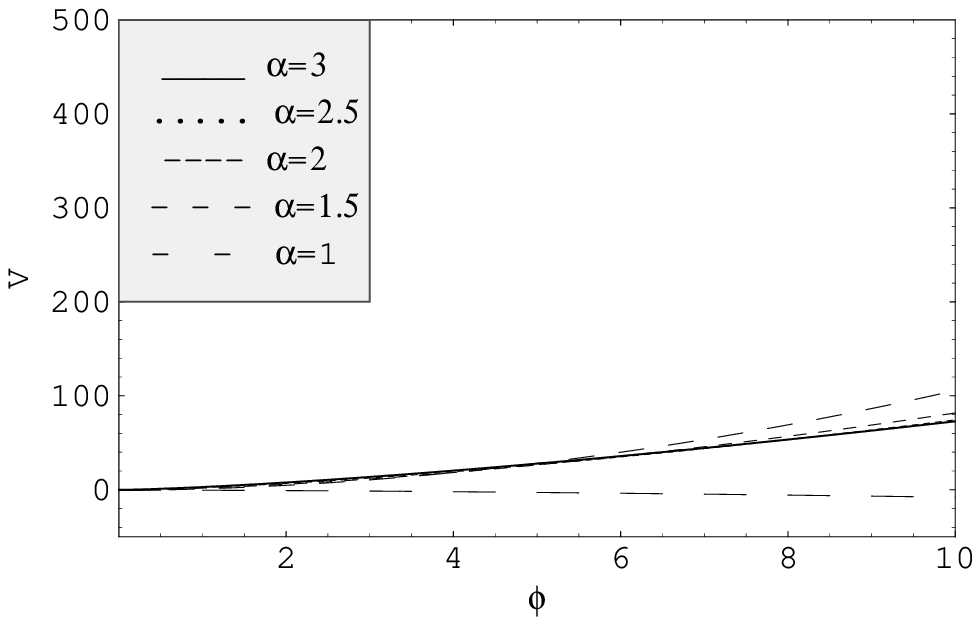}
\includegraphics[height=1.3in]{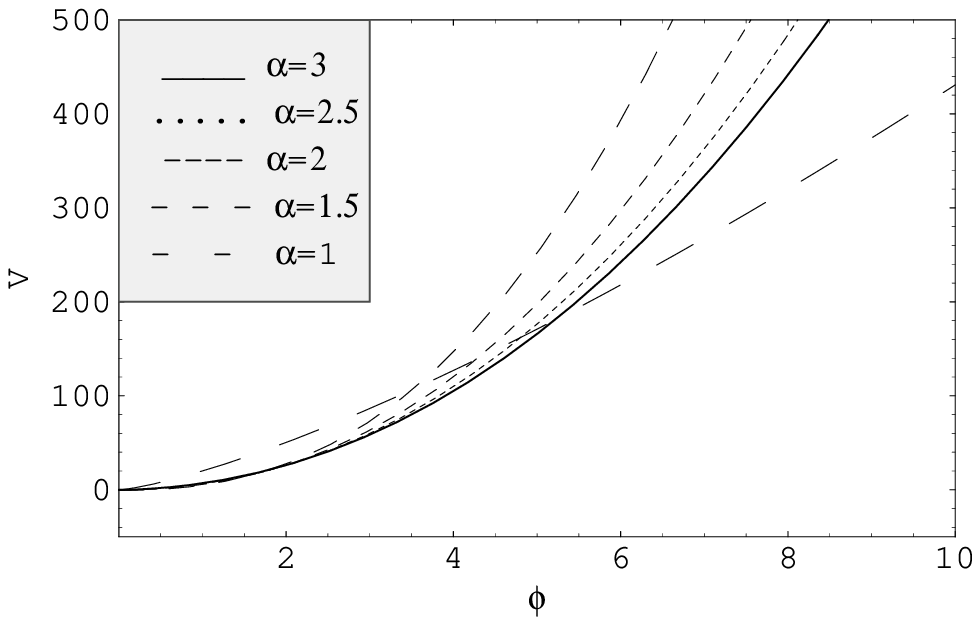}
\includegraphics[height=1.3in]{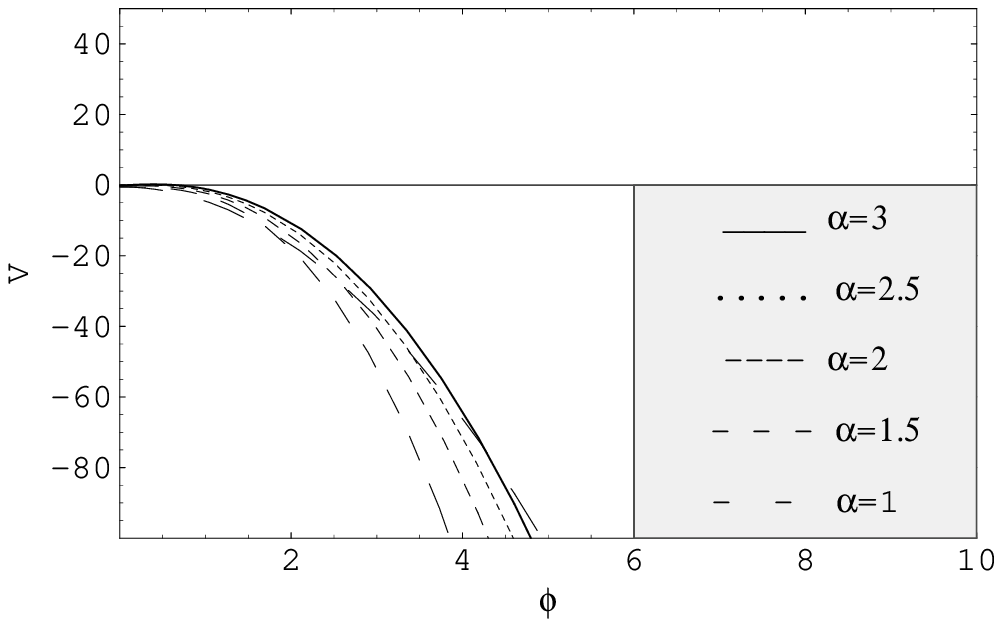}\\
\vspace{1mm}
Fig.16~~~~~~~~~~~~~~~~~~~~~~~~~~~~~~~~~~~~~~~~Fig.17~~~~~~~~~~~~~~~~~~~~~~~~~~~~~~~~~~~~~~~~~~~~~Fig.18\\
\vspace{2mm}Generalized Chaplygin Gas:\\
\vspace{2mm}
$V=V(\phi),\beta=-2\alpha,k=0$~~~~~~~~~~~~~~~~$V=V(\phi),\beta=-2\alpha,k=1$~~~~~~~~~~~~~~~~~~~~~~~~$V=V(\phi),\beta=-2\alpha,k=-1$
\vspace{5mm}

\vspace{5mm} Fig. 16, 17 and 18 shows the variation of $V$
 against $\phi$ for respectively flat, closed and open models of the Universe.
 We have considered different values of $\alpha=1, 1.5, 2, 2.5,3$ and $\beta=-2\alpha$.
 In all the three the figures we have considered
$n=1$,  normalizing the parameters as $a_{0}=\rho_{0}=\phi_{0}=B=C=1$. \hspace{14cm} \vspace{4mm}

\end{figure}

\begin{figure}
\includegraphics[height=1.2in]{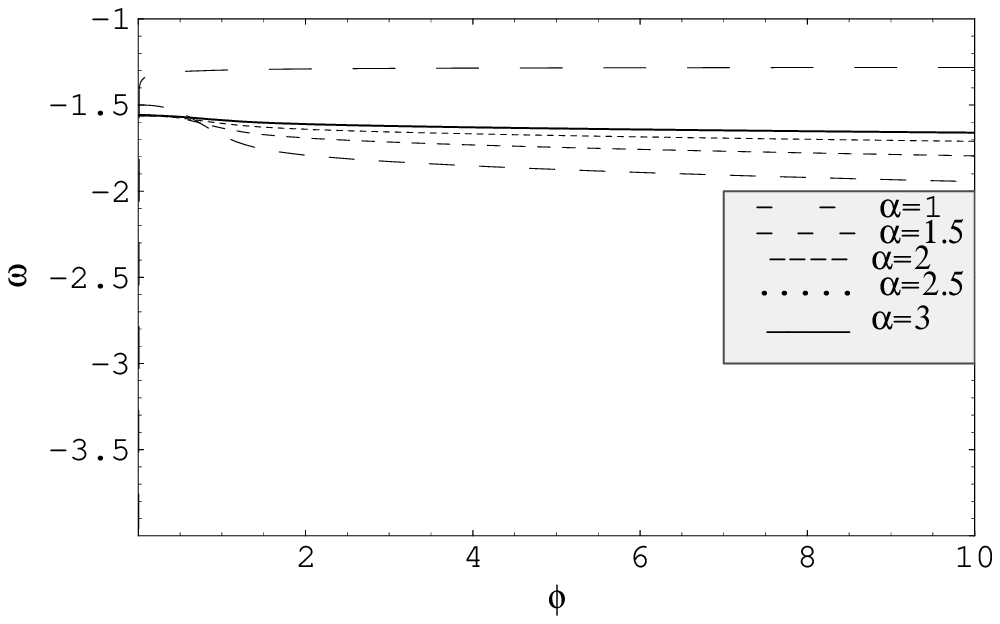}~~~~~~~~~~~~~~~~~~~~~~~~~~~~~~
\includegraphics[height=1.2in]{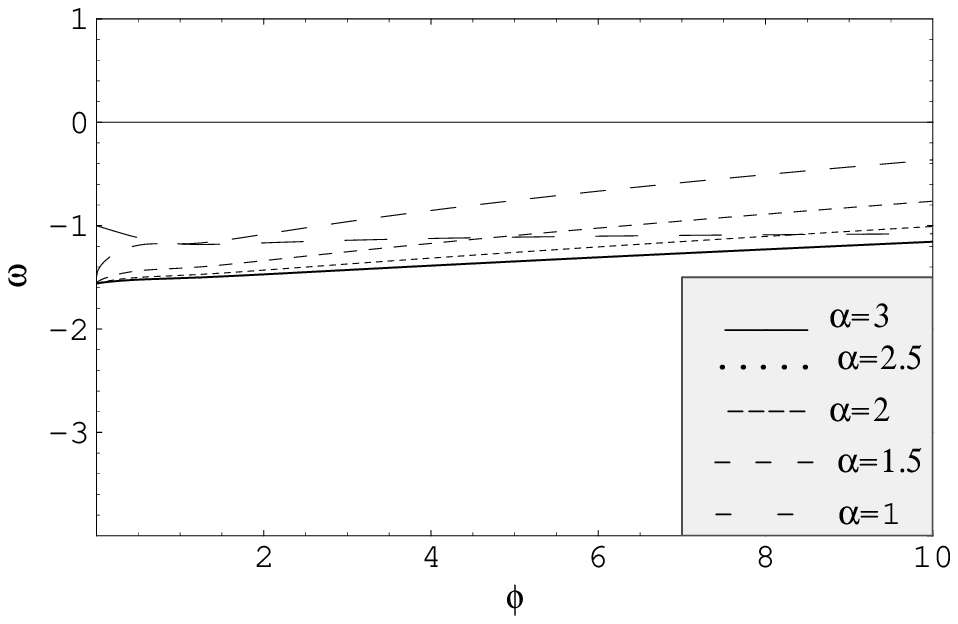}\\
\vspace{1mm}
Fig.19~~~~~~~~~~~~~~~~~~~~~~~~~~~~~~~~~~~~~~~~~~~~~~~~~~~~~Fig.20\\
\vspace{2mm}Generalized Chaplygin Gas:\\
 \vspace{2mm}
$\omega=\omega(\phi),k=0,\beta=-2\alpha$~~~~~~~~~~~~~~~~~~~~~~~~~~~~~~~~~~~~~~~~~~~
$\omega=\omega(\phi),k=1,\beta=-2\alpha$\vspace{5mm}

\vspace{5mm} Fig. 19 and 20 shows the variation of $\omega$
 against $\phi$ for different values of $\alpha=1, 1.5, 2, 2.5,3$.
 Here we have considered  $n=0$,
  normalizing the parameters as $a_{0}=\rho_{0}=\phi_{0}=b=C=1$,
  also as the calculation shows, for this $\beta=-2\alpha$. \hspace{14cm} \vspace{4mm}

\end{figure}

\section{\normalsize\bf{Conclusion}}

We are considering Friedman-Robertson-Walker model in Brans-Dicke
Theory with and without potential ($V$). Also we have considered
the Brans-Dicke parameter ($\omega$) to be constant and variable.
We take barotropic fluid
and Generalized Chaplygin Gas as the concerned fluid.\\

Using barotropic equation of state, we get, $(i)$ for $V=0$ and
$\omega=$constant, $\omega<0$ and $q<0$ for some values of
$\alpha$, giving rise to cosmic acceleration, $(ii)$ for $V=0$
and $\omega=\omega(\phi)$, we obtain cosmic acceleration
depending on some values of $\alpha$ and $\beta$. In this case we
get acceleration for closed model also at the radiation phase. We
can show the variation of $\omega(\phi)$ against the variation of
$\phi$ here [figure 1,2]. Figure 1 shows that as the value of
$\alpha$ increases $\omega$ decreases steadily against the
variation of $\phi$. For $\alpha>1$, we have accelerated
expansion. The figure shows that the greatest value of $\omega$
can be $-\frac{3}{2}$ and it decreases further as $\phi$
increases, $(iii)$ for $V=V(\phi)$ and $\omega=$constant, we get
acceleration in the flat model irrespective of the values of
$\alpha$, $(iv)$ for $V=V(\phi)$ and $\omega=\omega(\phi)$,
cosmic acceleration is obtained for $\beta\le 2$. Here we can
represent the variation of $\omega$ and $V$ against the variation
of $\phi$ for $\beta=-2$ and $\beta=-2\alpha$. For $\beta=-2$,
the variation of $\omega$ against $\phi$ is same as figure 1 and
that for closed and open models are given in figure 3 and 4. Here
we can see that for open model $\omega$ starting at
$-\frac{3}{2}$ decreases further, whereas for closed model
$\omega$ starting at $-\frac{3}{2}$ increases to e positive for
$\alpha=2$. Figures 5, 6 and 7 sow that variation of $V$ against
the variation of $\phi$ for $\beta=-2$ in respectively flat,
closed and open models of the Universe. Here we can see that only
for the closed model the potential increases positively, in the
other two cases the potential becomes negative after a certain
point. Figure 8 shows the variation of $V$ against $\phi$ for
$\beta=-2\alpha$. Again positive potential energy is obtained for
only the closed model. The variation of $\omega$ is shown in
figure 9 for $k=1$ and we can see that $\omega$ increases starting
at $-\frac{3}{2}$.\\

Using Generalized Chaplygin Gas , we get, $(i)$ for $V=0$ and
$\omega=$constant, uniform expansion is obtained, $(ii)$ for $V=0$
and $\omega=\omega(\phi)$, Generalized Chaplygin Gas does not
seem to have any effect of itself, $(iii)$ for $V=V(\phi)$ and
$\omega=$constant, we get $q=0$ giving uniform expansion, $(iv)$
for $V=V(\phi)$ and $\omega=\omega(\phi)$ cosmic acceleration is
obtained for $\beta\le 2$ as previously obtained for barotropic
fluid. Figures 10, 11, and 12 show the variation of $\omega$ for
$\beta=-2$ in flat closed and open models and the natures of the
graphs do not vary much from that for barotropic fluid. Figures
13, 14 and 15 show the variation of $V$ for flat, closed and open
models respectively. Here for open model we get a negative
potential after a certain point, whereas for closed model we get
a positive potential always. For spatially flat model a positive
$V$ is obtained for $\alpha=1.5, 2$. Figures 16, 17 and 18 show
the variation of $V$ for the models of the Universe for
$\beta=-2\alpha$. Positive potential is obtained for closed model
and flat model shows positive potential for $\alpha>1$. For open
model we get negative $V$ again. Figures 19 and 20 show the
variation of $\omega$ for flat and closed models respectively
($\beta=-2\alpha$). For flat model $\omega$ starting at
$-\frac{3}{2}$ decreases further and for closed model it
increases slowly from $-\frac{3}{2}$.\\

We have used B-D Theory to solve the problem of cosmic
acceleration. Here we use barotropic fluid and Generalized
Chaplygin Gas. Although the problem of fitting the value of
$\omega$ to the limits imposed by the solar system experiments
could not be solved fully, for closed Universe and $\beta=-2$ and
$\alpha>1$, $\omega$ starting from $-\frac{3}{2}$ increases and
for large $\phi$, we get $\omega>500$, for both barotropic fluid
and Generalized Chaplygin Gas. Also for flat Universe filled with
barotropic fluid taking $\omega=$constant and $V=V(\phi)$, we get
the Bertolami-Martins [11] solution, i.e, $V=V(\phi^{2})$ and
$q_{0}=-\frac{1}{4}$ for $a=At^\frac{4}{3}$. But taking
Generalized Chaplygin Gas, we get accelerated expansion only when
both $\omega$ and $V$ are functions of the scalar field $\phi$.
For $\beta=-2$ we get cosmic acceleration in the closed model,
whereas, $\beta=-2\alpha$ gives acceleration in both closed and
flat models of the Universe, although for flat Universe $\omega$
varies from $-\frac{3}{2}$ to $-2$ and for closed Universe
$\omega$ takes large values for large $\phi$. In the end we see
that for all the cases accelerated expansion can be achieved for
closed model of the Universe for large values of $\omega$. Also
the present day acceleration of the Universe can also be explained
successfully, although in this
case $\omega$ cannot meet the solar system limits.\\\\

{\bf Acknowledgement:}\\

The authors are thankful to IUCAA, India for warm hospitality
where part of the work was carried out. Also UD is thankful to
UGC, Govt. of India for providing research project grant (No. 32-157/2006(SR)).\\

{\bf References:}\\
\\
$[1]$ S. J. Perlmutter et al, {\it Bull. Am. Astron. Soc.} {\bf
29} 1351 (1997).\\
$[2]$ A. G. Riess et al, {\it Astron. J.} {\bf 116} 1009 (1998); B. P. Schmidt et al, {\it Astrophys. J.} {\bf 507} 46 (1998).\\
$[3]$ R. R. Caldwell, R. Dave and P. J. Steinhardt, {\it Phys. Rev. Lett.} {\bf 80} 1582 (1998).\\
$[4]$ A. S. Al-Rawaf and M. O. Taha, {\it Gen. Rel. Grav.} {\bf
28} 935 (1996).\\
$[5]$ T. Padmanabhan, {\it Phys. Rept.} {\bf 380} 235 (2003).\\
$[6]$ N. Banerjee and D. Pavon, {\it Class. Quantum Grav.} {\bf 18} 593-599 (2001).\\
$[7]$ L. P. Chimento, A. S. Jakubi and D. Pavon, {\it Phys. Rev. D} {\bf 62} 063508 (2000).\\
$[8]$ N. Banerjee and D. Pavon, {\it Phys. Rev. D} {\bf 63} 043504 (2001).\\
$[9]$ B. K. Sahoo and L. P. Singh, {\it Modern Phys. Lett. A} {\bf 18} 2725- 2734 (2003).\\
$[10]$ K. Nordtvedt,Jr., {\it Astrophys. J} {\bf 161} 1059 (1970); P. G. Bergmann, {\it Int. J. Phys.} {\bf 1} 25 (1968);
R. V. Wagoner, {\it Phys. Rev. D} {\bf 1} 3209 (1970).\\
$[11]$ O. Bertolami and P. J. Martins, {\it Phys. Rev. D} {\bf 61} 064007 (2000).\\
$[12]$ V. Gorini, A. Kamenshchik and U. Moschella, {\it Phys. Rev.
D} {\bf 67} 063509 (2003); U. Alam, V. Sahni , T. D. Saini and
A.A. Starobinsky, {\it Mon. Not. Roy. Astron. Soc.} {\bf 344}, 1057 (2003).\\
$[13]$ M. C. Bento, O. Bertolami and A. A. Sen, {\it Phys. Rev. D}
{66} 043507 (2002).\\

\end{document}